\documentclass[%
 aip,
 sd,%
 amsmath,amssymb,
 reprint,%
]{revtex4-1}

\usepackage{graphicx}
\usepackage{sidecap}
\usepackage{dblfloatfix}
\usepackage{dcolumn}
\usepackage{bm}
\newcommand\simlt{\lower.5ex\hbox{$\; \buildrel < \over \sim \;$}}
\newcommand\simgt{\lower.5ex\hbox{$\; \buildrel > \over \sim \;$}}

\begin{document}

\preprint{AIP/123-QED}

\title[]{Dynamics of Relativistic Shock Waves Subject to a Strong Radiation Drag: Similarity Solutions and Numerical Simulations}

\author{Ilia Leitus}
\author{Amir Levinson}%
 \email{Levinson@wise.tau.ac.il}
\affiliation{ Raymond and Beverly Sackler School of Physics \& Astronomy, Tel Aviv University, Tel Aviv 69978, Israel
}%


\date{\today}

\begin{abstract}
We examine the effect of Compton drag on the dynamics of a relativistic shock wave. 
Similarity solutions describing a radiation-supported shock are obtained for certain 
profiles of the external radiation intensity and the density of the unshocked ejecta, and
are compared with 1D numerical simulations of a blast wave expanding into an ambient 
medium containing isotropic seed radiation.  Both the analytic model and the simulations indicate 
that under realistic conditions the radiation drag should strongly affect the dynamics and structure of the shocked layer.
 In particular, our calculations show significant strengthening of the reverse shock and weakening of the forward shock 
 over time of the order of the inverse Compton cooling time.   We conclude that the effect of radiation drag on the 
 evolution of the emitting plasma should affect the resultant light curves and, conceivably,
spectra of the observed emission during strong blazar flares.

\end{abstract}

\maketitle

\section{\label{Introduction} Introduction}

High-energy emission is a defining property of compact relativistic astrophysical sources.   There is ample 
evidence that the emission is produced in relativistic outflows launched by  an accreting
black hole or a magnetar.    The interaction of these outflows with their environments results in formation 
of expanding cocoons and shocks, that decelerate the flow and lead to lower energy emissions over a  large range of scales,
e.g., afterglow emission in GRBs and radio lobes in AGNs and micro-quasars.    In certain circumstances the relativistic jets interact 
also with ambient radiation fields emitted by the surrounding gas.   This interaction can considerably alter the dynamics 
of dissipative fluid shells at relatively small scales,  and is likely to be the origin of the variable gamma-energy emission observed
in certain classes of high-energy sources.

Direct evidence for such interactions is found in blazars, where the properties of the ambient radiation field can be measured.
This radiation is contributed primarily by  the accretion disk around the black hole,  by gaseous clouds in the broad line 
region (BLR) and, at larger scales, by a dusty molecular torus \cite{JMB14,SSMNM09}.   The conventional wisdom has
been that  the high-energy emission seen in powerful  blazars 
results from inverse Compton scattering of ambient seed photons by non-thermal electrons accelerated 
in dissipative regions in the jet \cite{DS93,SBR94,BL95,GM96}. In most early works the dynamics of the radiating fronts is not computed in a self-consistent manner.  More recent works 
incorporated realistic shock models to compute light curves of flared emission \cite{JMB14,JB11}.
However, the effect of the radiation drag on the structure and dynamics of the shock has been neglected. 

In this paper we construct a model for the dynamics of a relativistic shock wave in the presence of intense radiation field. 
In section \ref{sec:II} we derive the governing equations of a radiation-supported shock.
In section  \ref{sec:self-similar}  we present self-similar solutions, obtained under several simplifying 
assumptions regarding the properties of the ejecta and the intensity profile of the external radiation field, 
and study the dependency of the evolved structure on the magnitude of the radiation drag.  In section  \ref{sec:sim} 
we present results of numerical simulations of a uniform, spherical ejecta propagating through an ambient medium 
containing isotropic seed radiation.  The simulations enable analysis of the temporal evolution from the 
initial stage under more realistic conditions.  While the analysis is motivated by the application to blazars, mainly 
because there we have sufficient information to asses the conditions in the source, it might also be relevant to 
other high-energy astrophysical systems.

\section{\label{sec:II}Radiation-supported shock model}

Quite generally, the collision of relativistic fluid shells leads to the formation 
of a double-shocked layer.  Some examples include the interaction of expanding ejecta with 
an ambient medium and collision of fast and slow shells (internal shocks).  
In the absence of external forces, the dynamics of the double-shock
system is dictated solely by the properties of the unshocked media.  In the presence of
external radiation field, the shocked layers are subject to radiation drag that 
can alter the dynamics significantly.  When the drag is strong enough it should
lead to a weakening of the forward shock and strengthening of the reverse shock.  In 
essence, the compression of the shocked ejecta (or the shocked plasma of the fast shell 
in case of internal shocks) is supported by the radiation pressure imposed on it.  
Thus, as a simple approximation one can ignore the entire region beyond the contact discontinuity, and
obtain solutions describing a radiation-supported shock.  This is the treatment adopted in section \ref{sec:self-similar} below.
The validity of this approximation is checked in section \ref{sec:sim} using numerical simulations.

Below we adopt the following notation:
The velocities of the reverse shock and the contact surface are denoted by $V_r$ and $V_c$, respectively,
and the corresponding Lorentz factors by $\Gamma_r$ and $\Gamma_c$.  Plasma quantities are denoted by 
lower case letters, and subscripts $u$ and $s$ refer to the unshocked and shocked ejecta, respectively.

\subsection{Flow equations}

The dynamics of a relativistic fluid is governed by the continuity equation,  
\begin{equation}\label{cont}
    \partial_{\mu}\left(\rho u^{\mu}\right)=0,
\end{equation}
and energy-momentum equations,
\begin{equation} \label{EMten}
    \frac{\partial}{\partial x^{\alpha}}T^{\alpha\mu}=S^\mu,
\end{equation}
here expressed in terms of the energy-momentum tensor
\begin{equation} \label{Tmunu}
T^{\mu \nu}=w\, u^\mu u^\nu + g^{\mu\nu} p,
\end{equation}
where $\rho$, $p$ and $w$ are, respectively, the proper density, pressure and specific enthalpy, 
$u^\mu=\gamma(1,v_1,v_2,v_3)$ 
is the fluid 4-velocity, and $g^{\mu\nu}$ is the metric tensor, here taken to be the Minikowsky metric
$g_{\mu\nu}={\rm diag}(-1,1,1,1)$.  The source term on the right hand side of Eq. (\ref{EMten}) accounts
for energy and momentum exchange between the fluid and the ambient radiation field, and is derived explicitly below.
The above set of equations must be augmented by an equation of state.  
We adopt the relation 
\begin{equation}
w=\rho+\frac{\hat{\gamma}}{\hat{\gamma}-1}p\equiv \rho+ap,
\end{equation}
with $a=4$ for a relativistically hot plasma and $a=5/2$ for sub-relativisric temperatures.

We shall restrict our analysis to spherical (or conical) flows, and adopt a spherical coordinate
system $(r,\theta,\phi)$.  Equations (\ref{cont}) and (\ref{EMten}) 
for the shocked fluid then reduce to:
\begin{eqnarray}
\partial_t(\rho\gamma)+\frac{1}{r^2}\partial_r(r^2\rho\gamma v)=0,\label{cont-radial}\\
\partial_t(w\gamma^2-p)+\frac{1}{r^2}\partial_r(r^2 w\gamma^2 v)=S^{0},\label{mot-radial0}\\
w\frac{d\ln\gamma}{dt}+\frac{dp}{dt}-\gamma^{-2}\partial_tp=v(S^r-v S^0),
\label{Eq-mot-radial}
\end{eqnarray}
here $d/dt=\partial_r +v\partial_r$ denotes the convective derivative.

A common approach is to seek solutions of Eqs. (\ref{cont-radial})-(\ref{Eq-mot-radial}) inside the 
shocked layer, given the properties 
of the unshocked flows upstream of the forward and reverse shocks as input.  The jump 
conditions at the forward and reverse shocks then serve as boundary conditions for
the shocked flow equations.  In addition, the fluid velocity and total momentum flux must 
be continuous across the contact discontinuity surface.  In the approximate treatment adopted here, 
whereby the reverse shock is fully supported by radiation and the layer enclosed between the contact and the 
forward shock can be ignored, the location of the contact is fixed by the requirement that the total energy is 
conserved, given formally by Eq. (\ref{Global_E_cons}) below.

\subsection{Jump conditions at the reverse shock}
For the range of conditions envisaged here the shocked plasma is optically thin.  The shock, in this case, is collisionless, i.e., 
mediated by collective plasma processes (as opposed to radiation mediated shocks), and its width, roughly the skin depth, is much smaller than 
the Thomson length.   Thus, the shock transition layer can be treated as a discontinuity in the flow parameters. 
The local jump conditions at the reverse shock can be expressed in terms of the local shock velocity $V_r=dR_r/dt$, where $R_r(t)$ is the 
shock radius at time $t$, as \cite{NS06,Le10}
\begin{eqnarray} 
&& \rho_{u}\gamma_{u}(v_u-V_r)=\rho_{s}\gamma_{s}(v_{s}-V_r), \label{basic_jump1}\\
&&  w_{u}\gamma_{u}^2v_u(v_u-V_r)+p_{u}=w_{s}\gamma_{s}^2v_s(v_{s}-V_r)+p_{s}, \label{basic_jump2}\\
&& w_{u}\gamma_{u}^2(v_u-V_r)+p_{u}V_r=w_{s}\gamma_{s}^2(v_{s}-V_r)+p_{s}V_r. \label{basic_jump3}
\end{eqnarray}   
We shall henceforth assume that the unshocked medium is cold and set $p_u=0$.
The jump conditions at the forward shock can, in principle, be derived in a similar manner, but are redundant under our radiation-supported shock approximation.

\subsection{Compton drag terms}
Behind the shock electrons are heated and accelerated to relativistic energies.  
We denote the energy distribution of electrons there by $F_e(\gamma_e)$, where 
$\gamma_e=\epsilon/m_ec^2$ is the dimensionless electron energy, as measured in the fluid frame.  The total electron 
proper density is then given by $n_e=\int F_e(\gamma_e) d\gamma_e$, and the average energy and second energy moment by 
\begin{equation}
<\gamma_e>=\frac{1}{n_e}\int \gamma_e  F_e(\gamma_e) d\gamma_e,\quad <\gamma_e^2>=\frac{1}{n_e}\int \gamma_e^2  F_e(\gamma_e) d\gamma_e.
\end{equation}
If only the thermal electrons are taken into account one expects  $m_e<\gamma_e>\simeq m_p\gamma^\prime_u/2$ for plasma 
in rough equipartition, where 
$\gamma^\prime_u=\gamma_u\Gamma_r(1-v_uV_r)$ is the Lorentz factor of the upstream flow, as measured in the shock frame, and it is assumed that $\gamma^\prime_u>1$.  If the contribution of non-thermal electrons is substantial, then $<\gamma_e>$ may be larger. 
Since the shocked electron plasma is relativistic we have $m_ec^2n_e<\gamma_e^2>=w<\gamma_e>$.

The derivation of the source terms is given in Ref \onlinecite{GL15}.   In terms of the energy density of the 
radiation intercepted by the flow, $u_{rad}(r)$, and the quantities defined above they can be expressed as 
\begin{equation}
S^0=-\frac{8}{3}\gamma_s^3<\gamma_{e}^2>u_{rad}\sigma_T n_{e},
\label{S0c}
\end{equation}
and 
\begin{equation}\label{Src}
S^r=v_s S^{0}+S^0/3\gamma_s^2.  
\end{equation}
From Eqs. (\ref{Eq-mot-radial}) and (\ref{Src}) it is readily seen that the deceleration time 
of the shocked fluid is given by 
$t_{dec}=-3w\gamma_s^2/v_s S^0=2 \gamma_s t_{IC}^\prime$, where $t^\prime_{IC}\simeq 3m_ec/(4\sigma_T u^\prime_{rad}<\gamma_e>)$
is the inverse Compton cooling time of an electron having an energy $m_ec^2<\gamma_e>$, as measured in the fluid rest frame, and 
$u_{rad}^\prime=4\gamma_s^2 u_{rad}/3$.  Substantial deceleration occurs if $t_{dec}$ is much shorter than the outflow time, $t_f=r/c$.
For convenience, we define the drag coefficient as
\begin{equation}
\alpha=3t_f/t_{dec}=\frac{8}{3 m_e c^2}\gamma_s<\gamma_e> \sigma_T u_{rad} r.
\end{equation}
As noted above, for thermal electrons $\gamma_s<\gamma_e>\simeq (m_p/2m_e)\gamma^\prime_u\gamma_s
=(m_p/4m_e)\gamma_u$.  If only the contribution of thermal electrons is accounted for, then
\begin{equation}\label{drag_thermal}
\alpha=\frac{8 m_p \sigma_T}{12 m^2_e c^2}\gamma_u u_{rad}\, r.
\end{equation}
The energy loss term can now be expressed as:
\begin{equation} \label{sc0-2}
 S^0=-\alpha\gamma_s^2 w r^{-1}.
\end{equation}

To get an estimate of the value of $\alpha$ anticipated in blazars, suppose that a fraction $\eta$ of the nuclear luminosity, $L=10^{45} L_{45}$
erg/s, is scattered across the jet.  The corresponding energy density of the radiation intercepted by the jet is roughly 
$u_{rad}(r)=\eta L/4\pi r^2=3\times10^{-4}\eta L_{45}/r_{pc}^2$ ergs 
cm$^{-3}$, where $r_{pc}$ is the radius in parsecs.   From Eq. (\ref{drag_thermal}) one obtains $\alpha= 0.7 \gamma_u\eta  L_{45} r_{pc}^{-1}$. 
A more realistic estimate can be obtained using recent calculations of 
the ambient radiation intensity contributed by the various radiation sources surrounding a blazar jet \cite{JMB14}.  
It is found that for a prototypical blazar the intensity profile of radiation intercepted by the jet is flat up to a distance of about one parsec, 
with $u_{rad}\simeq 10^{-3}$ erg cm$^{-3}$, followed by a gradual decline .   For an internal shock 
produced by colliding shells, Equation (\ref{drag_thermal}) yields for the drag coefficient $\alpha\approx 3\gamma_u r_{pc}$.   
Typically $\gamma_u >10$, implying significant drag on sub-parsec scales in blazars.

\subsection{Global energy conservation}
The change over time in the total energy of the shocked ejecta (or shocked fast shell) must equal the difference between the energy injected through 
the reverse shock and the energy radiated away through the contact.   The energy accumulated behind the forward shock is neglected in our
approximate treatment. 
The time change of the total energy per steradian of the shocked ejecta is
\begin{eqnarray}
\partial_t E &=& \partial_t \int_{R_r(t)}^{R_c(t)}T^{00}r^2 dr = \int_{R_r(t)}^{R_c(t)}\partial_t(T^{00}r^2) dr \nonumber \\
&+&R_c^2 T^{00}_s(R_c)V_c-R_r^2 T^{00}_s (R_r)V_s,\label{d_tE}
\end{eqnarray}
where $R_r(t)$ and $R_c(t)$ are, respectively, the radii of the reverse shock and the contact surface, and 
$V_{r} =dR_{r}/dt$, $V_c=dR_c/dt$ are the corresponding velocities. 
Integrating Eq. (\ref{mot-radial0}) from the shock, $R_r(t)$, to the contact, $R_c(t)$, one obtains:

\begin{eqnarray}
& & \int_{R_r(t)}^{R_c(t)} \partial_t(T^{00}r^2) dr +R_c^2 w_s(R_c)\gamma_s^2(R_c) v_s(R_c)\nonumber  \\
& -& R_r^2 w_s(R_r)\gamma_s^2(R_r) v_s(R_r)=\int_{R_r(t)}^{R_c(t)}S^0 r^2 dr.\label{energy_balance}
\end{eqnarray}
Combining Eqs. (\ref{d_tE}) and (\ref{energy_balance}), using the jump condition  (\ref{basic_jump3}) with $p_u=0$ (cold ejecta) and the relation $T^{00}=w \gamma^2 -p$, and recalling that the fluid velocity is continuous across the contact, viz., $v_s(R_c)=V_c$, yields
\begin{equation}\label{Global_E_cons}
\partial_t E =R_r^2 \rho_u\gamma_u^2(v_u-V_r)-R_c^2p_s(R_c)V_c +\int_{R_r(t)}^{R_c(t)}S^0 r^2 dr.
\end{equation}
The first term on the right hand side accounts for the power incident through the reverse shock, the second term for $pdV$
work at the contact and the last term for radiative losses.

\section{\label{sec:self-similar}Self-similar solutions}
Similarity solutions describing the interaction of a relativistic shell with an ambient medium, in the absence 
of radiative losses, 
were derived in Ref \onlinecite{NS06}, and their stability was subsequently analyzed \cite{Le09,Le10}.    Such solutions can be obtained for a
freely expanding ejecta characterized by a velocity profile $v_u=r/t$ at time $t$ after the explosion, and  
a proper density of the form
\begin{equation}
\rho_u=\frac{a_u}{t^3\gamma_u^n}, \label{den-ejecta}
\end{equation}
where $a_u$ is a normalization constant, and $\gamma_u=1/\sqrt{1-v_u^2}$ is the corresponding Lorentz factor (it can be 
readily seen that the continuity equation is satisfied for this choice of $\rho_u$ and $v_u$).   
Self-similarity requires that the Lorentz factors of the forward shock, reverse shock and the contact discontinuity have a similar 
time evolution, viz., $\Gamma^2_f=At^{-m}$, $\Gamma^2_r=Bt^{-m}$, $\Gamma^2_c=Ct^{-m}$, where 
$A,B,C$ and $m$ are constants determined upon matching 
the solutions in region 1 (shocked ejecta) and region 2 (shocked ambient medium) at the contact discontinuity\cite{NS06,Le10}.  
For an ambient density profile of the form $\rho_i\propto r^{-k}$, the index $m$ is given by \cite{NS06,Le10}
\begin{equation}\label{m-value}
m=\frac{6-2k}{n+2}.
\end{equation}

In cases where the shocked plasma is subject to strong radiative losses it is still possible to obtain self-similar solutions provided
the energy source term scales as $S^0\propto \gamma_s^2 w r^{-1}$ (see Eq. (\ref{mot-radial0}) and (\ref{Eq-mot-radial})), that is, 
$\alpha$ in Eq. (\ref{sc0-2}) is constant.   We shall henceforth make this assumption even though it implies a somewhat artificial intensity
profile of the external radiation field.  We note, however, that as long as the deceleration length is smaller than the radius of the shock these
details are unimportant. This is confirmed by numerical simulations, presented in the next section.

As explained above, in the presence of a strong radiation drag the dynamics of the system is dictated by the interaction of the external radiation
field with the shocked plasma enclosed between the contact and the reverse shock.  To a good approximation one can then ignore the
contribution of the forward shock to the overall dynamics.  The evolution of the radiation-supported shock still satisfies  $\Gamma^2_r=Bt^{-m}$, 
$\Gamma^2_c=Ct^{-m}$, but the index $m$ is now determined from global energy conservation, as will be shown below.   Now, 
to order $O(\Gamma_r^{-2})$ the trajectory 
of the reverse shock is given by
\begin{equation}
R_r(t)=\int_0^t{\left(1-\frac{1}{2\Gamma_r^2}\right){\rm d}t^\prime}=t-\frac{t}{2(m+1)\Gamma_r^2},\label{r2}
\end{equation}
from which we obtain for the velocity of the ejecta crossing the shock: $v_u(R_r)=R_r/t=1-1/[2(m+1)\Gamma_r^2]$.
The corresponding Lorentz factor is thus given, to the same order, by
\begin{eqnarray}\label{gamma_u/Gamma_r}
\gamma_u^2=(m+1)\Gamma_r^2,
\end{eqnarray}
and the density profile by
\begin{eqnarray}
\rho_u= \frac{a_u}{(m+1)^{n/2} B^{3/m}}\Gamma_r^{(6/m-n)}.
\end{eqnarray}
We adopt the similarity parameter introduced in Ref \onlinecite{BMK76},
\begin{equation}
\chi=[1+2(m+1)\Gamma_r^2](1-r/t).
\end{equation}
With this choice the reverse shock is located at $\chi=1$, and the contact at $\chi_c=\Gamma^2_r/\Gamma_c^2=B/C<1$.
Following Ref.~\onlinecite{NS06} we define the self-similar variables, $G(\chi)$, $H(\chi)$ and $F(\chi)$, such that at the reverse shock they satisfy the boundary conditions $G(1)=H(1)=F(1)=1$.  Upon solving the jump conditions (\ref{basic_jump1})-(\ref{basic_jump3}), the shocked fluid quantities
can be expressed in terms of the self-similar variables as

\begin{eqnarray}
& &\gamma_s^2=q\Gamma_r^2G(\chi),\\
& &\rho_s\gamma_s=\frac{m q \rho_u\gamma_u}{(m+1)(q-1)}  H(\chi),\\
& &p_s=\frac{m\rho_u}{a(q-1)+2}(1-\sqrt{q/(m+1)}) F(\chi),\\
& & w_s=\rho_s+ap_s\equiv K(\chi)p_s,
\end{eqnarray}
where the parameter $\sqrt{q}$ is the only positive root of the polynomial equation
\begin{equation}
\hat{\gamma}x^3+(2-\hat{\gamma})\sqrt{m+1}\,x^2-(2-\hat{\gamma})x-\hat{\gamma}\sqrt{m+1}\,x=0.\label{q}
\end{equation}
Upon substituting these relations into Eqs. (\ref{cont-radial})-(\ref{Eq-mot-radial}), and using Eq. (\ref{sc0-2}) with a constant drag coefficient, $\alpha=$ const \footnote{More generally, $\alpha$ can be taken to be any function of the similarity coordinate $\chi$. In the presence of strong drag the width of the shocked layer is much smaller the its radius, hence we anticipate little variations in $\alpha$ across it.}, we obtain the following set of equations for the self-similar variables:
\begin{eqnarray}
2(1&+&qG\chi)\partial_\chi \ln F-(1-qG\chi)K\partial_\chi \ln G\nonumber \\
&=&\frac{mn-6-(m-\frac{2}{3}\alpha)K}{(m+1)}qG, \label{self-1}\\ 
2(1&-&qG\chi)\partial_\chi \ln F-\hat{\gamma}(1+qG\chi)\partial_\chi\ln G\nonumber \\
&=&\frac{6-mn+(m-4)\hat{\gamma}-\frac{2}{3}(\hat{\gamma}-1)
\alpha K}{m+1}qG,\label{self-2} \\
2(1&-&qG\chi)\partial_\chi \ln H - 2\partial_\chi\ln G\nonumber\\
&=&-\frac{(mn-m-2)}{(m+1)}qG,\label{self-3}
\end{eqnarray}
subject to the boundary conditions $G(1)=F(1)=H(1)=1$.  These equations reduce to those derived in 
Ref.~\onlinecite{NS06} in the special case $\alpha=0$.  
As can be seen, at the contact, $\chi=\chi_c$, the Lorentz factor must satisfy $qG_c\chi_c=1$, where for short we denote $G(\chi_c)=G_c$. 
This relation defines the limit of integration.  

The self-similar equations involve two eigenvalues; the index $m$ and the location of the contact $\chi_c$.   Thus, two conditions are needed
to find them. The first one is the relation 
\begin{equation}\label{eigen-1}
q G_c\chi_c=1 
\end{equation}
mentioned above.   The second one is global energy conservation, Eq. (\ref{Global_E_cons}).
To order $O(\gamma_s^{-2})$ we have $T_s^{00}=w_s\gamma_s^2-p_s\simeq w_s\gamma_s^2$,  $r=t$ and $dr=-t d\chi/[2(m+1)\Gamma_r^2]$, 
from which we obtain
\begin{eqnarray}
E(t)&=&\int_{R_r(t)}^{R_c(t)}T^{00}r^2dr = \frac{t^{n m/2} q m a_u (1-\sqrt{q/(m+1)})}{2(m+1)^{n/2+1}B^{n/2}[a(q-1)+2]}\nonumber \\
&\times&\int_{\chi_c}^1K(\chi)F(\chi)G(\chi) d\chi,
\end{eqnarray}
and $\partial_t E=nm E/2t$.  To the same order one has $\int_{R_r}^{R_c}S^0r^2dr = -\alpha E/t$.  Substituting these results into Eq. (\ref{Global_E_cons}) yields the constraint 
\begin{eqnarray}
\int_{\chi_c}^1K(\chi)&F&(\chi)G(\chi) d\chi=\frac{(m+1)}{q(\alpha+nm/2)}\nonumber \\
&\times& \left[ \frac{a(q-1)+2}{1-\sqrt{q/(m+1)}}-2F_c  \right],\label{eigen-2}
\end{eqnarray}
where $F_c=F(\chi_c)$.

\subsection{Results}
\begin{figure}[h!]
\centering
\centerline{\includegraphics[width=8cm]{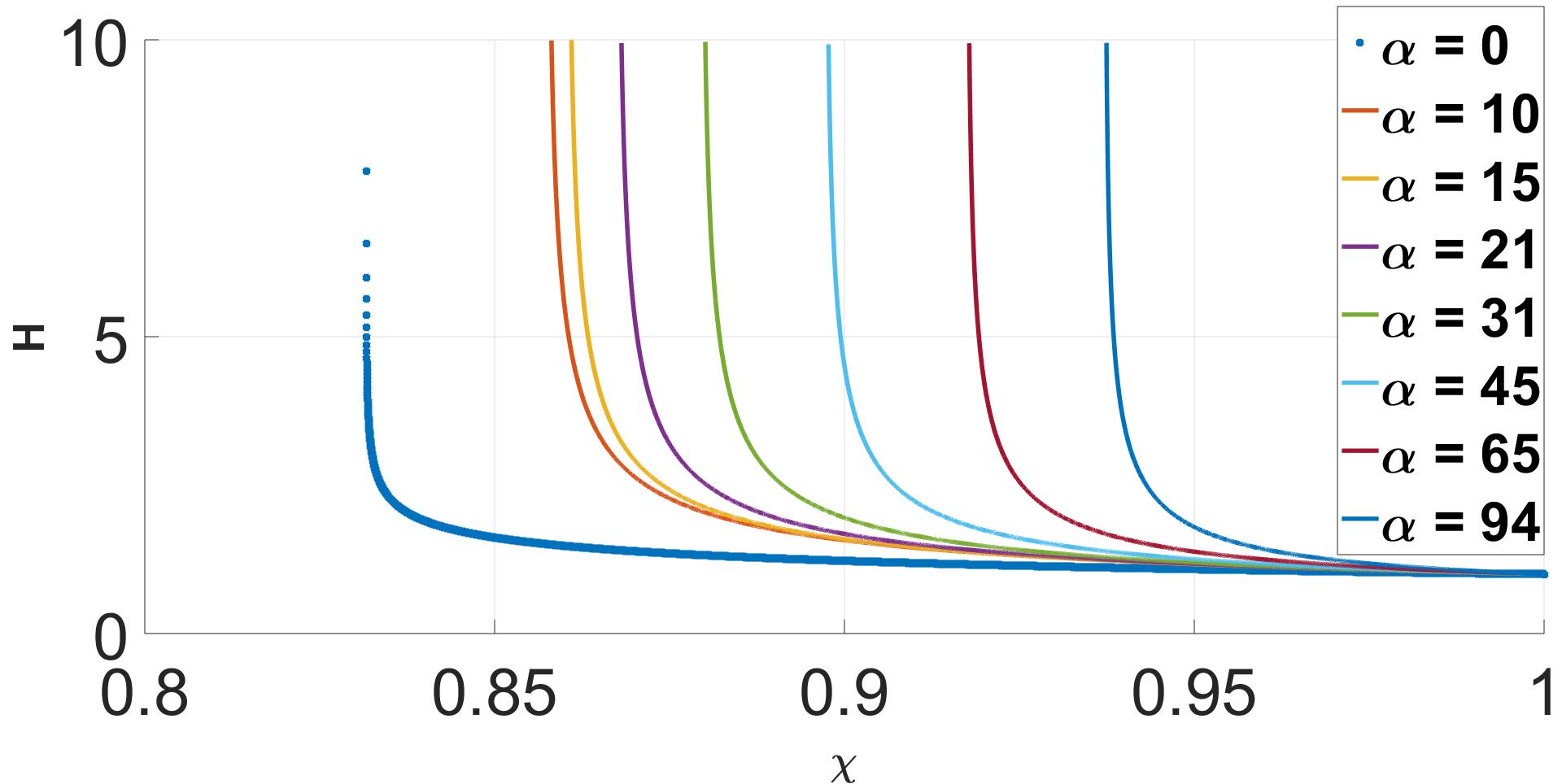}}
\centerline{ \includegraphics[width=8cm]{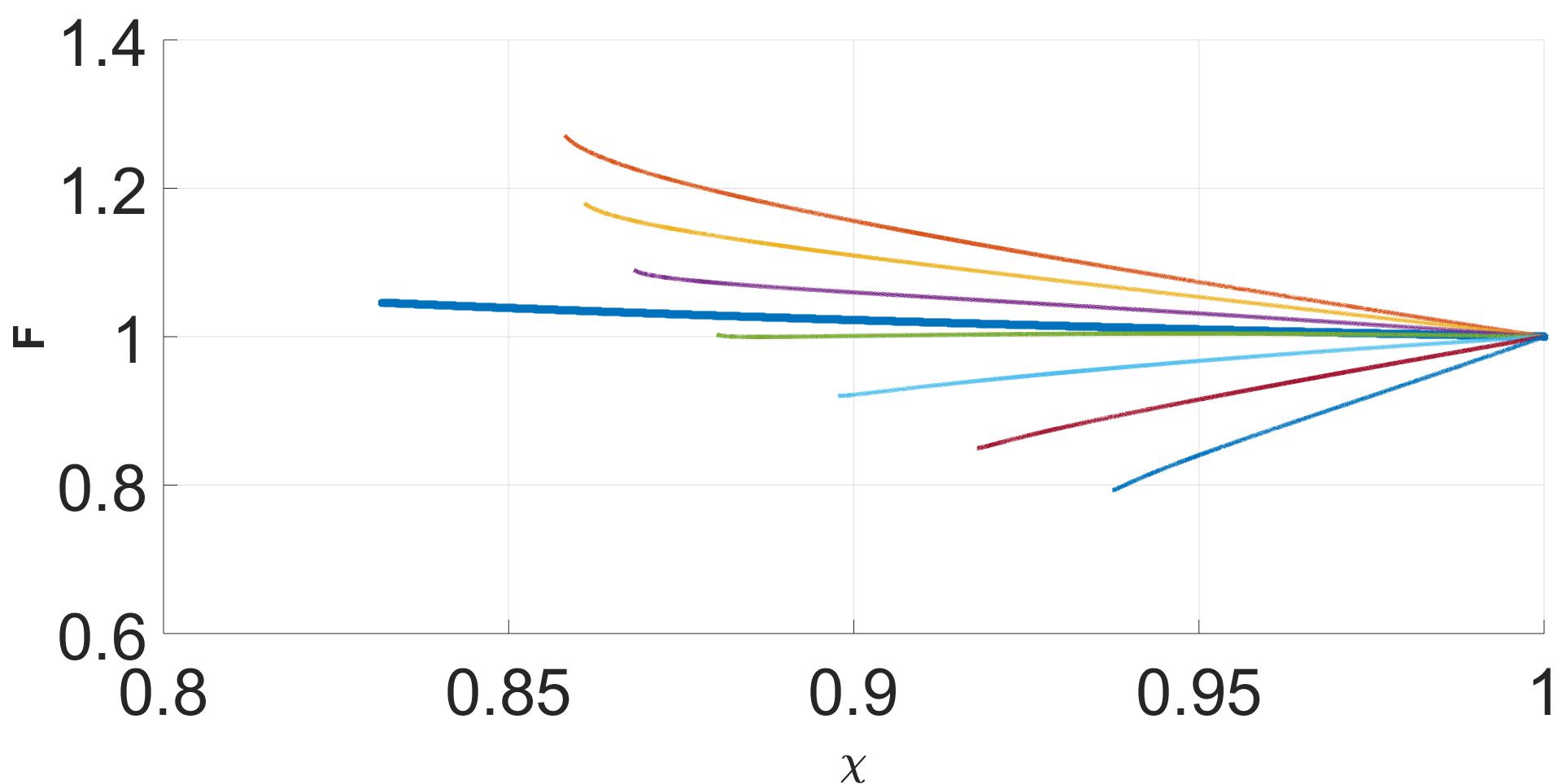}}
\centerline{\includegraphics[width=8cm]{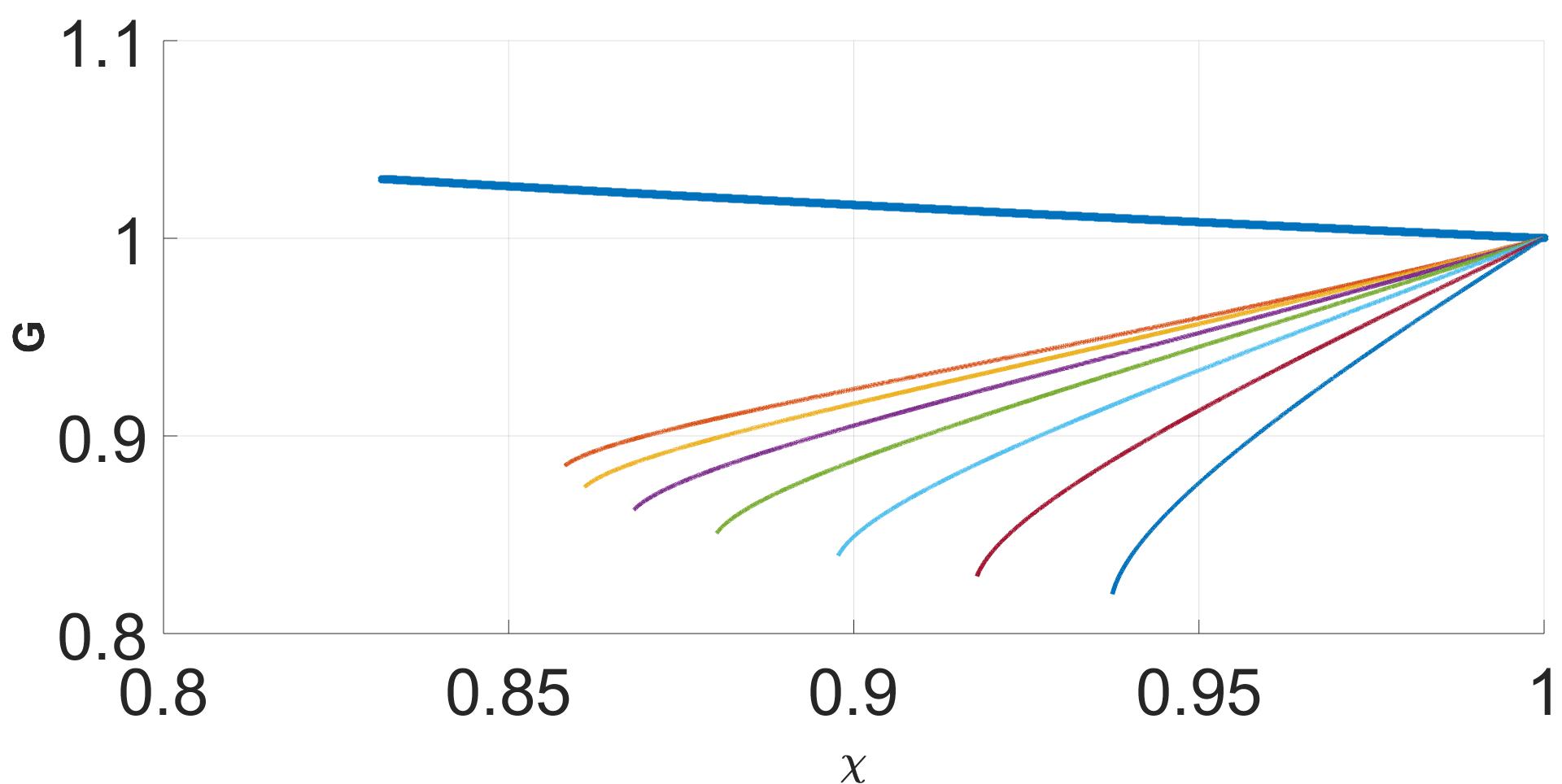}}
\caption{Profiles of the normalized density (top panel), pressure  (middle panel) and Lorentz factor (bottom panle), plotted as functions of the 
self-similar parameter $\chi$, for different values of the drag coefficient $\alpha$, as indicated in the upper panel.}
\label{fig:f1}
\end{figure}

For a given choice of the drag coefficient $\alpha$ we guess the value of $m$ and integrate Eqs. (\ref{self-1})-(\ref{self-3}) from $\chi=1$ to the point $\chi_c$ at which Eq. (\ref{eigen-1}) is satisfied.   We then check if Eq. (\ref{eigen-2}) is satisfied.  If not, we change the value of $m$ and repeat the process until
a solution satisfying all constraints is found. 
Sample profiles are shown in Fig. \ref{fig:f1}  for $n=1$ and different values of $\alpha$.  The radiation free case ($\alpha=0$) is shown for a comparison.  It was computed using the full solution of the two-shock model described in Ref. \onlinecite{NS06}.  As seen, the width of the shocked layer, $\Delta\chi=1-\chi_c$, decreases
with increasing drag coefficient $\alpha$, as naively expected.   Moreover, larger radiative losses lead to increased non-uniformity  of the
Lorentz factor and pressure in the shocked layer.  The divergence of the density at the contact 
is a basic feature of the similarity solutions \cite{Le10} and occurs even in the absence of radiative losses ($\alpha=0$), as seen in the upper panel of Fig. \ref{fig:f1}.

\begin{figure}[h]
\centerline{\includegraphics[width=8.5cm]{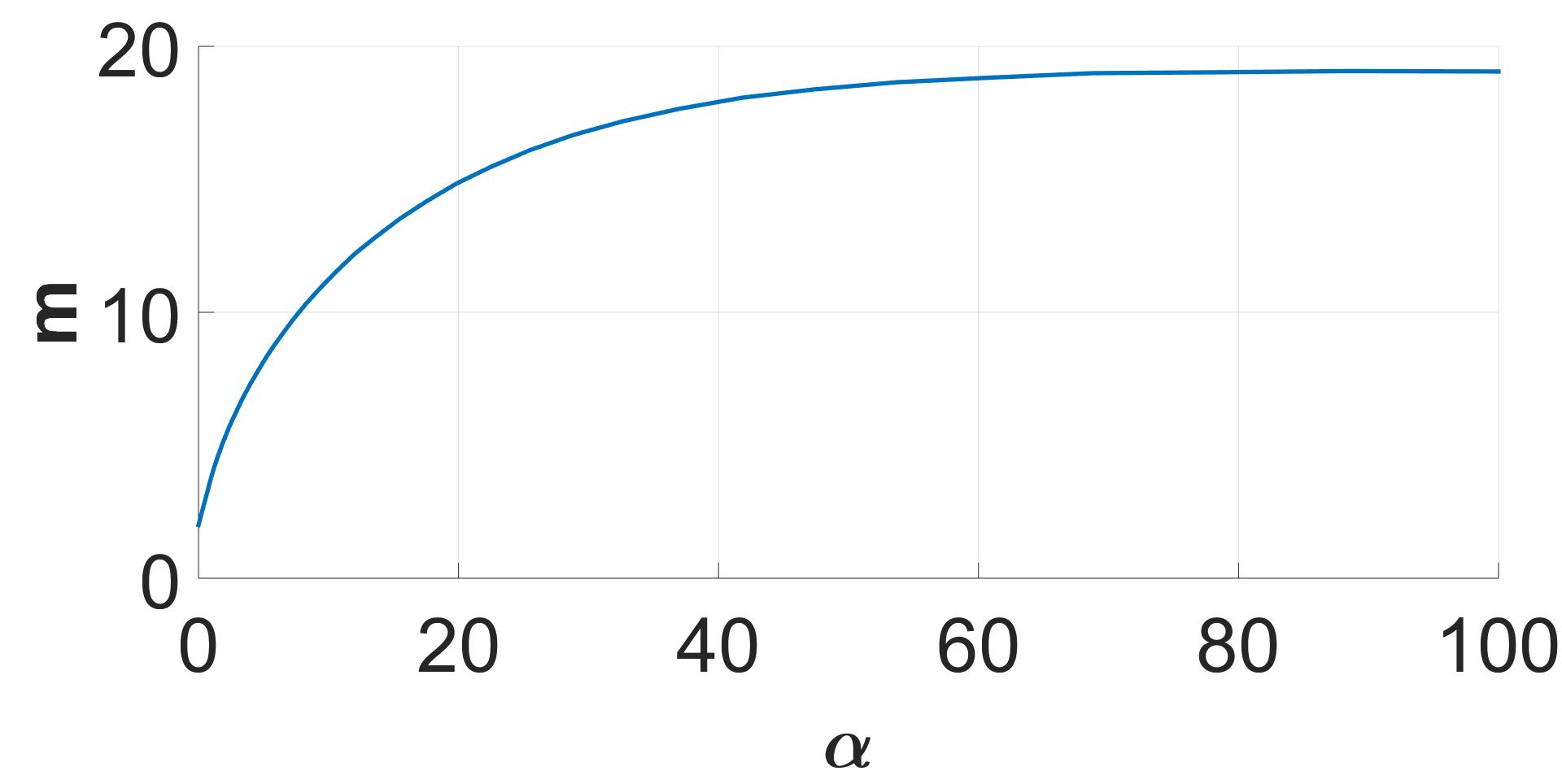}}
 \caption{Dependence of the index $m$ on $\alpha$}.
\label{fig:m}
\end{figure}

Figure \ref{fig:m} depicts the dependence  of the index $m$ on $\alpha$.   For a comparison, the value of $m$ obtained in the case $\alpha=0$ for a blast wave propagating in a uniform density medium ($k=0$) is $m=2$ (see Eq. (\ref{m-value}) with $k=0$ and $n=1$).   
The 4-velocity of the unshocked fluid, as measured in the frame of the reverse shock,  is given by
\begin{equation}
u_u^\prime=\gamma_u\Gamma_r(v_u-V_r)=\frac{m}{2\sqrt{m+1}},
\end{equation}
and it is seen that the shock becomes substantially stronger as $\alpha$ increases.   
Note that the power dissipated behind the shock, $\rho_u\gamma^\prime_uu^\prime_u$, increases roughly linearly with the index $m$.
At sufficiently large drag the entire power incident through the shock is radiated away, and the solution becomes independent of $\alpha$,
as seen in Fig. \ref{fig:m}.  For our choice of parameters, specifically $n=1$, this occurs at $\alpha\simgt 50$, for which $m\simeq 19$ and $u^\prime_u\simeq2.1$.    We emphasize that this limit can be approached provided the Lorentz factor $\gamma_u$ of the unshocked shell is sufficiently large.   
To be more concrete, Eq. (\ref{gamma_u/Gamma_r}) implies that $\gamma_u\simeq 4.6\Gamma_r$ as $\alpha\rightarrow \infty$, while our analysis is valid only for $\Gamma_r>>1$.   The Compton drag terms given in Eqs. (\ref{S0c}) and (\ref{Src}) assume that the intensity of ambient radiation is highly 
beamed in the frame of the shocked fluid.   Once $\gamma_s$ decelerates to modest values, $\gamma_s\simgt1$, the 
drag force is strongly reduced, ultimately becoming ineffective. 

\section{\label{sec:sim}Numerical simulations}
Numerical simulations were performed using the PLUTO code \cite{PLUTO07}, that was modified to incorporate energy and momentum losses of the shocked plasma through Compton scattering.  To clarify the presentation we adopted an external intensity profile of the form $u_s\propto r^{-1}$ for which $\alpha$ in Eq. (\ref{sc0-2}) is constant.   Over the deceleration scale this profile is a reasonable approximation to the flat profile expected in blazars \cite{JMB14}.
We start with the basic spherical blast wave problem, releasing an ejecta with a uniform Lorentz factor $\gamma_e = 10$ into a stationary ambient medium
(henceforth, quantities in the unshocked ejecta are designated by subscript "e").
In our setup, both the ejecta and the ambient medium are taken to be uniform initially with a density ratio $\rho_e/\rho_i = 100$. 
The calculations are performed in "simulation units" in which a fluid element traveling with a unit velocity (the speed of light) passes a unit of length per 
unit of time. The initial impact occurs at a radius $R_0 = 10^3$ and time $t = 0$.  
Due to the collision a double-shock structure forms and a shocked layer is created. As the simulation progresses the shocked layer widens and 
Compton drag is then applied on the shocked fluid contained between the forward and the reverse shocks. 
In order to suppress artificial transients created by abrupt changes we apply the radiation drag gradually by increasing the drag coefficient over
time as $\alpha(t) = \alpha (1- e^{-t/t_0})$.  As our model assumes that the drag only affects the hot shocked plasma, the in-simulation drag is applied
only in the region where the shocked fluid velocity is in the range $[0.1u_e, 0.9u_e]$, where $u_e$ is the 4-velocity of the unshocked ejecta. 
Preliminary  runs have shown that for the radiation free case the shocked layer becomes sufficiently developed by $t=30$ (see Fig. \ref{fig:test}), which is why the time constant for applying the radiation drag was chosen to be $t_0=10$ (by $t=30$ the drag reaches its maximum). For sufficiently high values of $\alpha$, the shocked layer decelerates and the reverse shock quickly becomes radiation-supported.

\subsection{Test case} \label{test_case}
\begin{figure}[h!]
    \centering
    \includegraphics[width=0.5\textwidth]{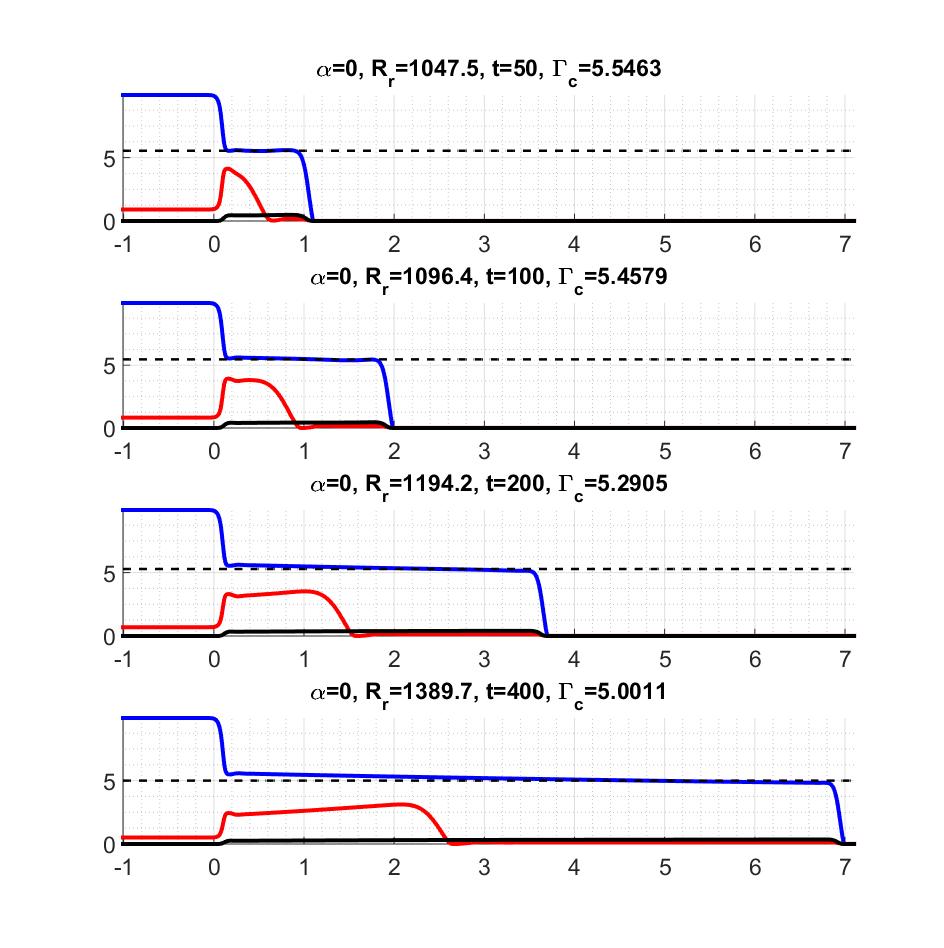}
    \caption{Profiles of the 4-velocity (blue solid line), pressure (black solid line), and density (red solid line) obtained from the test run
    of a radiation free system ($\alpha=0$), plotted as a functions of the distance from the reverse shock, $x=R-R_r(t)$. The value of the 
    contact 4-velocity $U_c$, computed from the analytic model, is marked by the 
    dashed line. As seen, the agreement is excellent.}
    \label{fig:test}
\end{figure}

As a check, we ran a test case with $\alpha=0$ for the same setup described above, and compared the results with an analytic solution
obtained under the thin shell approximation, whereby the shocked layers are assumed to be uniform.   Under this approximation
the shocked ejecta and shocked ambient medium have the same Lorentz factor, $\gamma_{s}=\Gamma_c$.  
In terms of the ratios  $q_e=(\gamma_e/\Gamma_r)^2$ and $q=(\gamma_{s}/\Gamma_r)^2$,
the jump conditions, Eqs. (\ref{basic_jump1}) - (\ref{basic_jump3}), yield
\begin{equation}
    2q_e\left(1-\sqrt{\frac{q}{q_e}}\right)+(a(q-1)+2)\left(\frac{q-q_e}{q+1}\right)=0,
\end{equation}
and
\begin{equation}\label{test-eq2}
    \frac{q}{q_e}=\frac{3}{4\gamma_e^2}\left(\frac{\rho_e}{\rho_i}\right)\frac{(q_e-1)\left(\sqrt{\frac{q}{q_e}}\right)}{q(q-1)+2},
\end{equation}
where $\rho_e$, $q$ and $q_e$ are functions of time.  
Solving these equations for the given initial conditions we obtain $q_e=3.765$, $q=1.2$ at $t=0$.  As the ejecta expands, its density just upstream 
of the reverse shock evolves as $\rho_e(t)\propto [R_r(t)]^{-2}$. Solving the above equations at any given time $t$ using $\rho_e(t)$ in Eq. (\ref{test-eq2}), 
one obtains the Lorentz factor of the contact discontinuity, $\Gamma_c(t)=\gamma_e\sqrt{\frac{q(t)}{q_e(t)}}$.  The contact 4-velocity,
$U_c(t)=\sqrt{\Gamma_c^2(t)-1}$, is shown as a dashed line in Fig.~\ref{fig:test}, along with the 4-velocity, pressure and density 
obtained from the simulations of the radiation free case ($\alpha=0$), at different times. 
As seen from Fig. \ref{fig:test}, the jump at the shock agrees well with the analytic result.  

\subsection{Simulation Results}
For numerical reasons we were only able to run cases with modest values of $\alpha$, however these suffice to illustrate the main trends.
Below we present results  for $ \alpha =3, 6, 10, 15$.   The initial conditions in all cases were set as described above.

\begin{figure}[h!]
    \centering
    \includegraphics[width=0.5\textwidth]{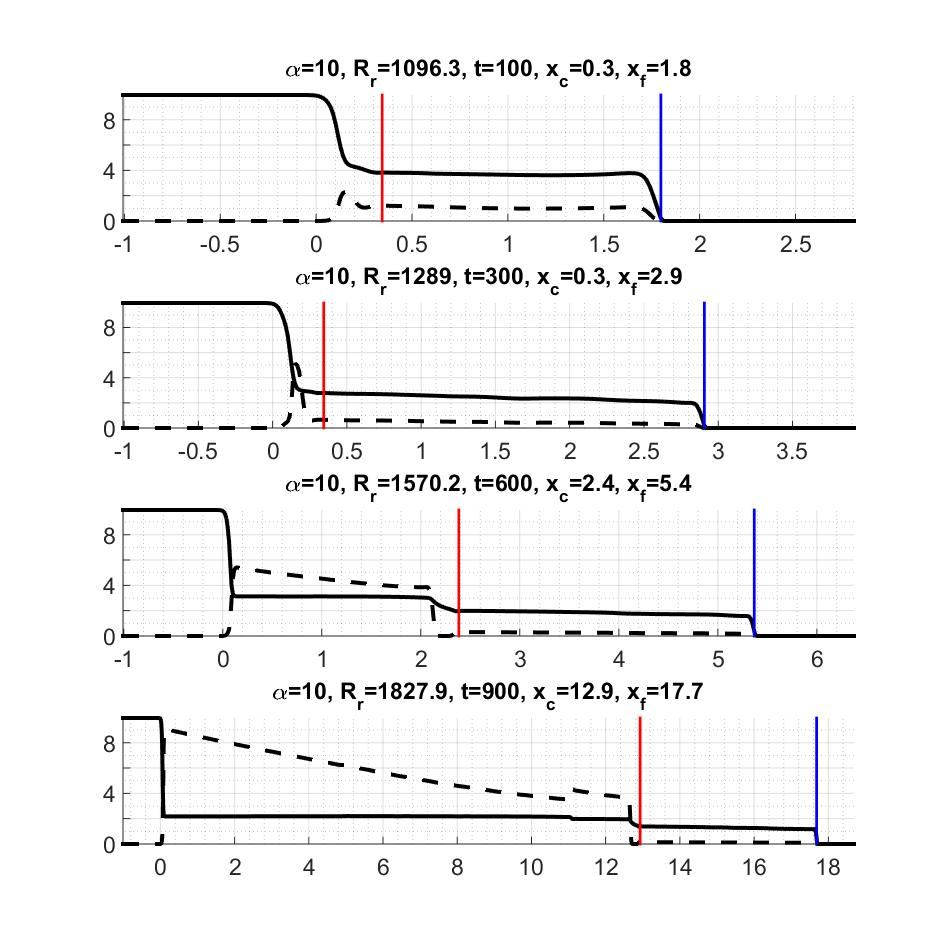}
    \caption{Profiles of the plasma 4-velocity (solid line) and pressure (dashed line) at different times, for the case $\alpha=10$.  
    For clarity of presentation, the pressure is scaled by a factor of $5$. Time progresses from top to bottom panels, as indicated. }
    \label{fig:a10_time}
\end{figure}

\begin{figure}[h!]
    \centering
 \includegraphics[width=0.5\textwidth]{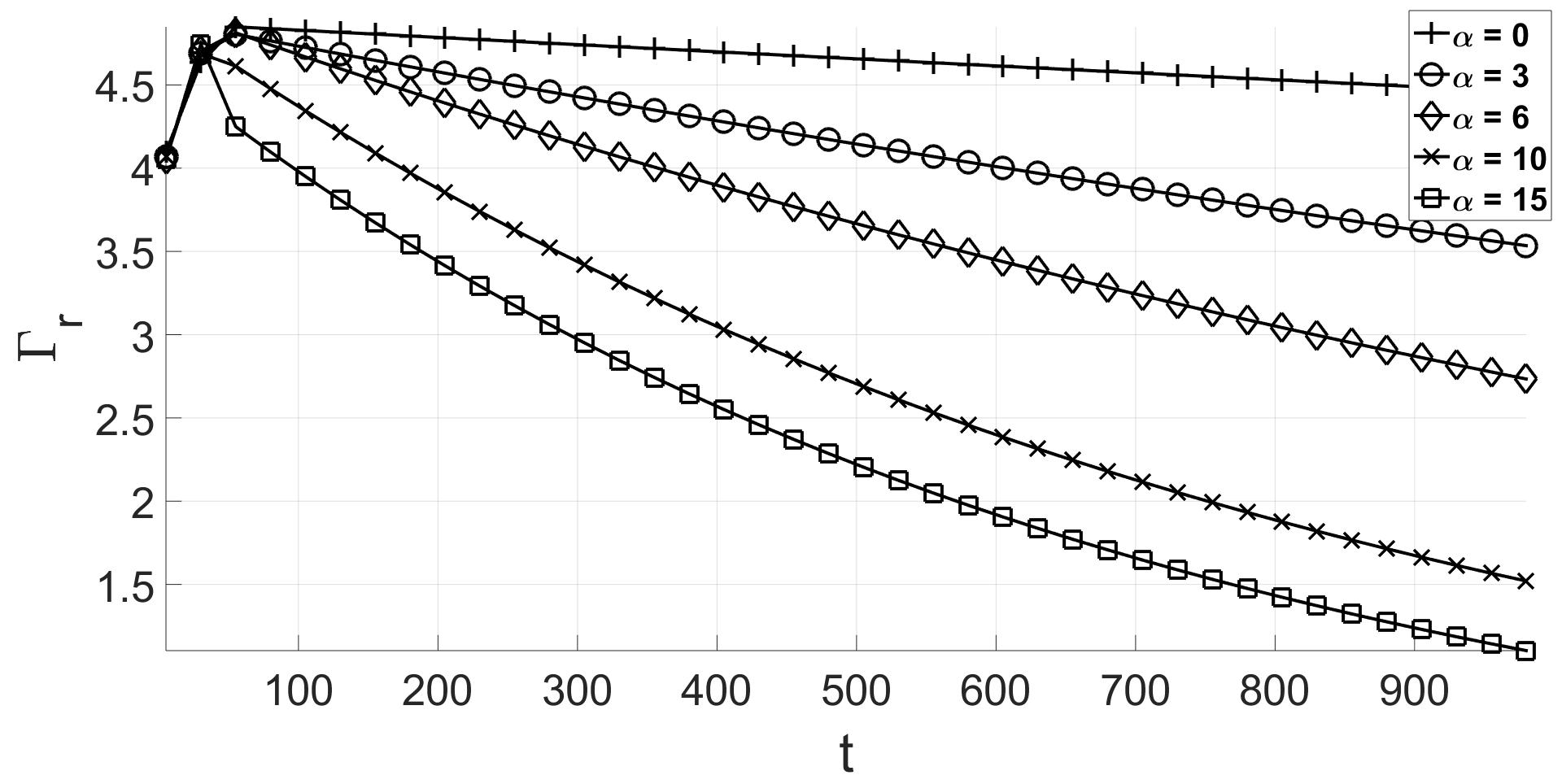}  
    \caption{Time evolution of the reverse shock Lorentz factor. Up to $t\approx30$ the shock structure is not sufficiently developed and the
    determination of $\Gamma_r$ is uncertain.}
    \label{fig:ur_t}
\end{figure}

Fig~\ref{fig:a10_time} displays snapshots of the plasma 4-velocity at different times for $\alpha=10$, showing the evolution of the entire 
structure.   The $x$-axis gives the distance from the reverse shock, $x=R-R_r(t)$, so that the reverse shock is located 
at $x=0$ at all times.   The vertical red and blue
lines indicate the location of the contact discontinuity $x_c$ and the forward shock $x_f$, respectively.  
 The deceleration of the shocked fluid, that leads to strengthening of the reverse shock and weakening of the forward shock is clearly
seen.   The radiation force gives rise to compression of the fluid near the contact which is
communicated  to the reverse shock and decelerates it (Fig. \ref{fig:ur_t}).   
The shocked layer gradually expands as time progresses, but initially less than in
the case $\alpha=0$.  The apparent sudden expansion of the shocked layer at late times (between $t=600$ and $t=900$) commences
when the shock becomes mildly relativistic, as can be observed from Fig.~\ref{fig:ur_t}.   This is mainly due to the fact that the width of the
shocked layer evolves with time as $t/\Gamma_r^2$.  We note that at such small Lorentz 
factors ($\Gamma_r<2.5$ at $t>600$) our choice of Compton drag terms, that assume perfect beaming of the external radiation in the 
rest frame of the shocked fluid, overestimates
the actual drag.  In reality the drag is expected to be suppressed as the shocked fluid becomes mildly relativistic , so that the Lorentz factor
of the reverse shock will eventually saturate once $\Gamma_r$ becomes sufficiently low.

\begin{figure*}[hb]
    \centering
\centerline{\includegraphics[width=8cm]{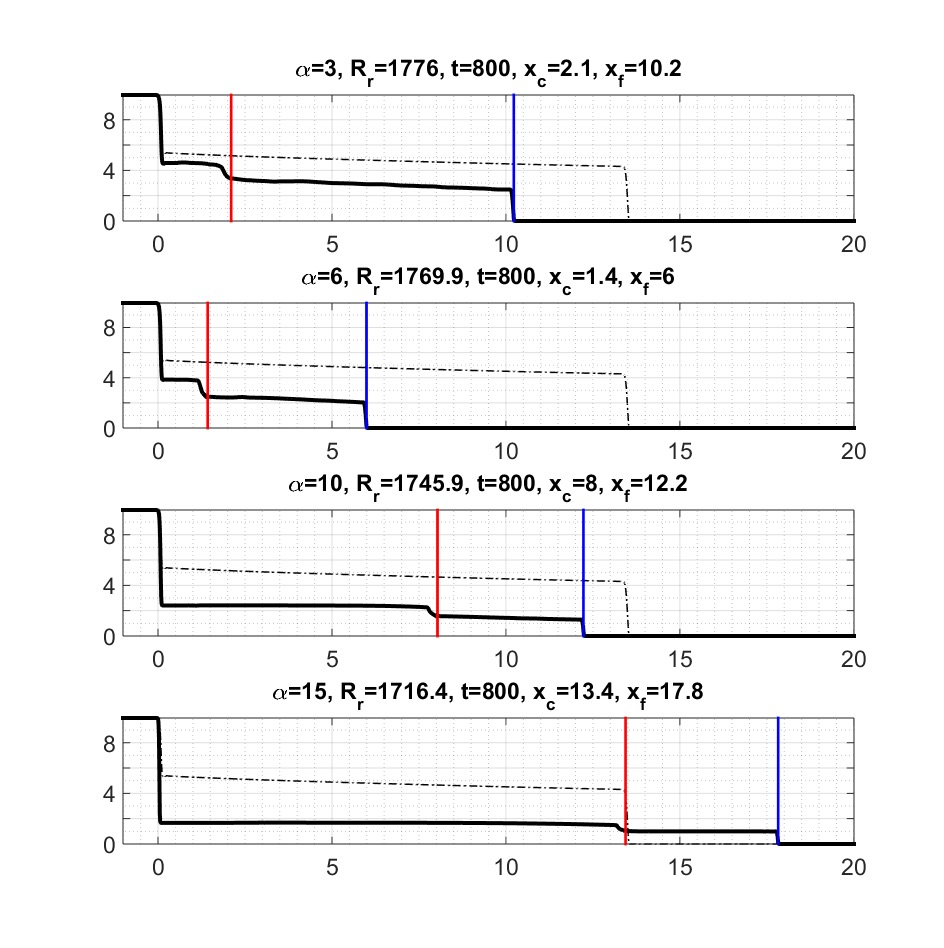} }
\centerline{
\includegraphics[width=8cm]{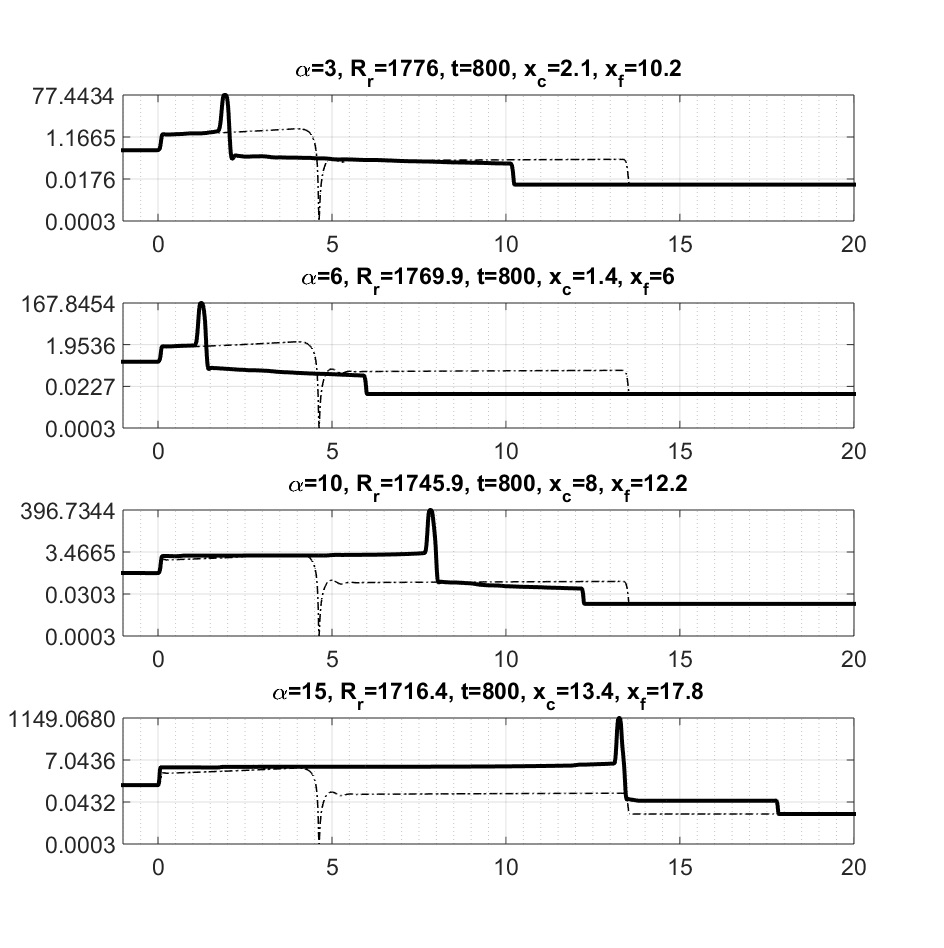} \includegraphics[width=8cm]{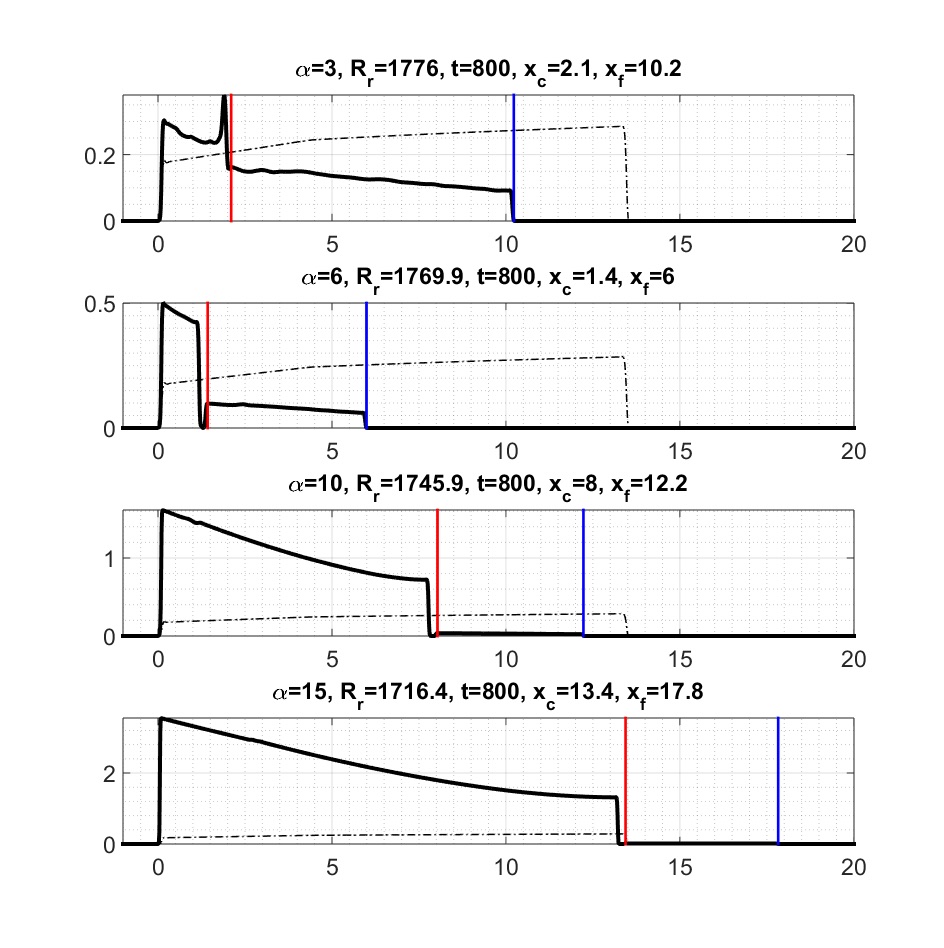}}
    \caption{Profiles of the 4-velocity (upper panel), density (lower left panel) and pressure (lower right panel) at simulation time $t=800$, plotted as a functions of the distance from the reverse shock, for different values of the drag coefficient $\alpha$.  The location of the reverse shock is at $x=0$.  The locations of the contact and the forward shock are marked by red and blue vertical lines respectively.  The thin dotted-dashed lines correspond to the case $\alpha=0$, and are shown for comparison.  As seen, the strengthening of the reverse shock and weakening of the forward shock become more substantial as $\alpha$ increases.  Strong compression by the radiation force is also seen in the pressure and density plots.  Note the logarithmic scale of the y-axis in the density plots.}
    \label{fig:u_u_s}
\end{figure*}

Figs~\ref{fig:u_u_s} shows the profiles of the 4-velocity, pressure and density around the shock at  the same simulation time, $t=800$, for four values of $\alpha$ 
as indicated in the figure label.  The radius of  the reverse shock at this time is also indicated for convenience, and it ranges from $R_r\approx1776$ (or $(R_r-R_0)/R_0 \simeq 0.77$) for $\alpha=3$ to $R_r\approx1716$ for $\alpha=15$.  The result of the run with no radiation drag ($\alpha=0$), presented in Fig. \ref{fig:test}, is plotted as a dotted-dashed line and is exhibited for a comparison. 
The contact location is found from the density profile, noticing that, as in the radiation-free case, the density peaks towards the contact and drops to 
a minimum right after it.
In the resulting structure we see the transition from "material - supported" to  "radiation-supported" shock; as $\alpha$ is increased the forward shock becomes progressively weaker and ultimately negligible, while the reverse shock strengthens.  This trend justifies the neglect of the shocked ambient pressure in the 
self-similar solution outlined in section \ref{sec:self-similar}. 
 Fig~\ref{fig:u_u_s} confirms that, as long as the shock is sufficiently relativistic, the width of the shocked layer shrinks and the shocked fluid velocity decreases with increasing radiation drag, in accord with the self-similar solution.  
 The widening of the shocked layer for the $\alpha=10$ and $\alpha=15$ cases stems from the transition to the mildly relativistic regime, as explained above.   
The increase of the pressure with increasing $\alpha$ further indicates strong compression by the radiation force. 
The formation of the cold dense shell near the contact, seen in the lower left panel of Fig.~\ref{fig:u_u_s}, results from the
rapid cooling of the compressed plasma. 

\begin{figure*}[h]
    \centering
\centerline{  \includegraphics[width=8.5cm]{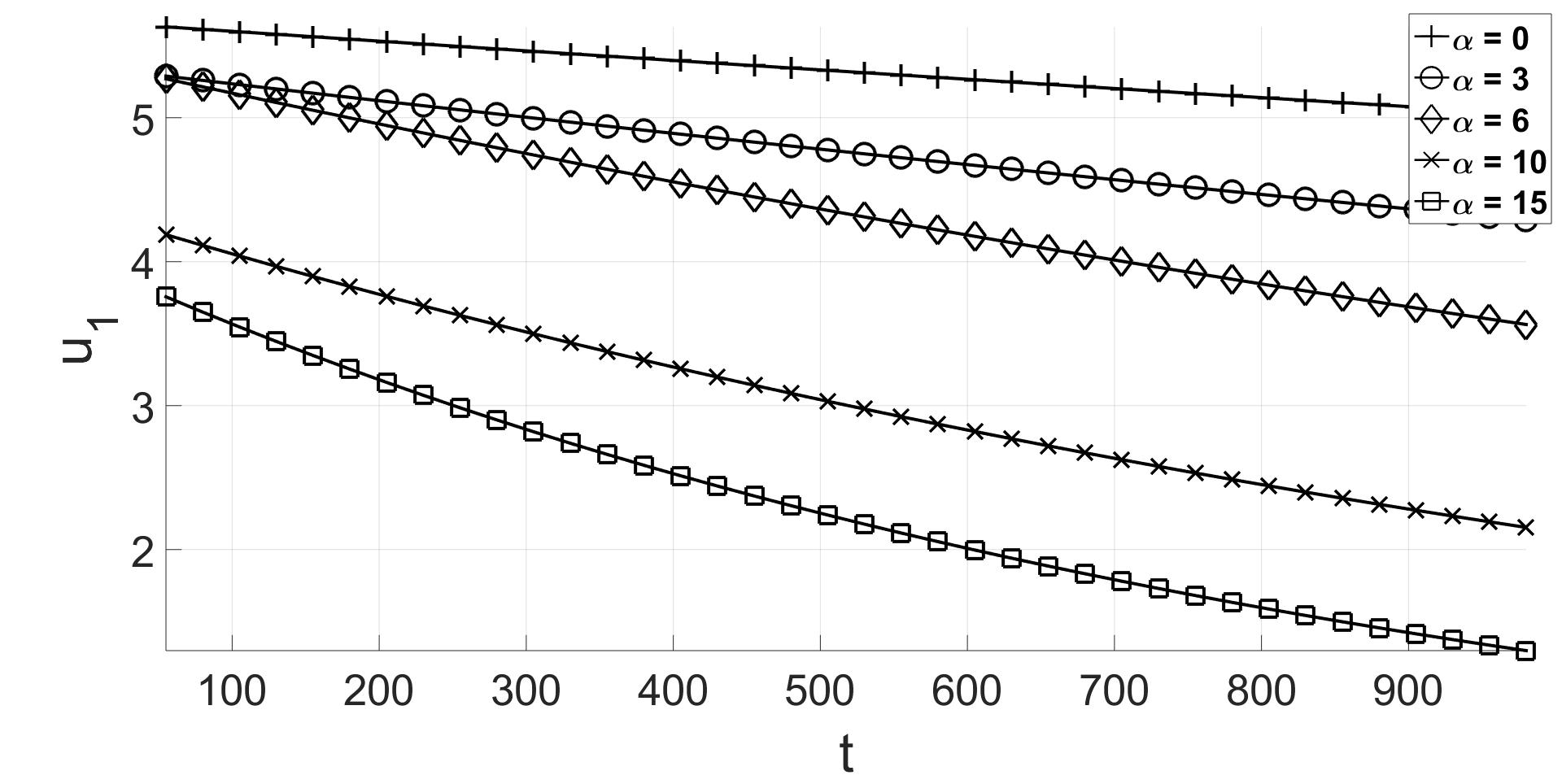} \includegraphics[width=8.5cm]{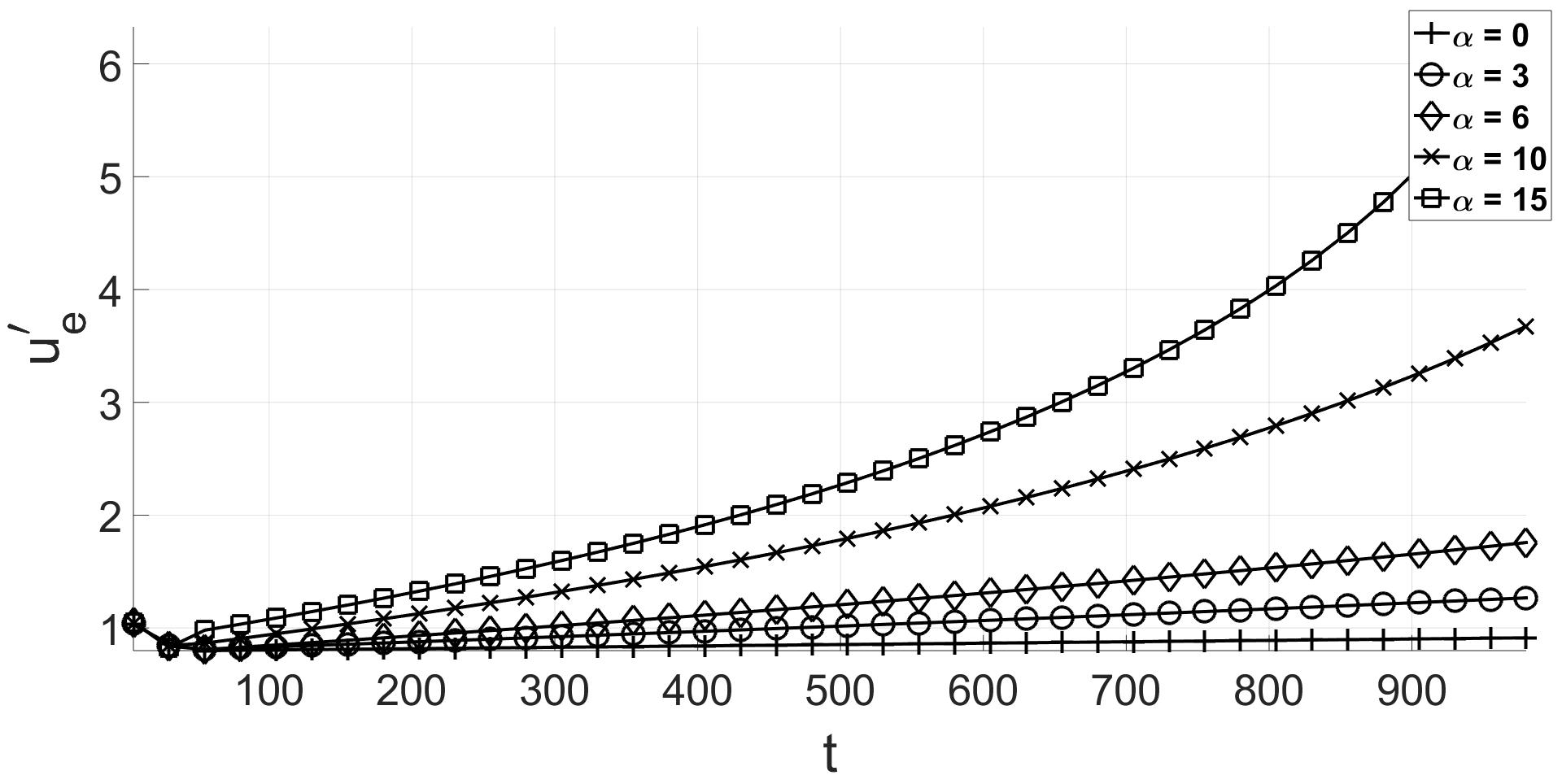}}
    \caption{Time evolution of the 4-velocity of the shocked ejecta (left panel), and of the unshocked ejecta, as measured in the shock frame (right panel).}
    \label{fig:u1_t}
\end{figure*}

The left panel of figure \ref{fig:u1_t} exhibits the evolution of the 4-velocity of the shocked ejecta.  Gradual deceleration of the shocked flow over a timescale of the order of the cooling time is observed, as expected.   The evolution of the 4-velocity of the unshocked ejecta with respect to the reverse shock is 
exhibited in the right panel of figure \ref{fig:u1_t}, indicating a substantial increase in shock efficiency with increasing $\alpha$. 

\section{Summary and conclusions}

We studied the effect of radiation drag on the dynamics of shocks that form in relativistic outflows. Such situations are expected in cases where the outflow propagates through a quasi-isotropic, ambient radiation field, on scales at which the inverse Compton cooling time is significantly shorter than the outflow expansion time. The observations of the continuum emission in blazars suggest that these conditions may prevail on sub-parsec to parsec scales in those objects. 

For certain profiles of the external radiation intensity and the density of the unshocked ejecta we were able to find self-similar solutions 
of the radiation hydrodynamics equations, describing a radiation-supported shock.   
We also performed 1D numerical simulations of a uniform, spherical shell interacting with an ambient medium that contains cold gas and seed radiation.  For that purpose we used the PLUTO code, that we modified to incorporate energy and momentum losses of the shocked plasma through Compton scattering. 

In both, the analytical model and the simulation results, we find significant alteration of the shock dynamics when the ratio of dynamical time and Compton cooling time exceeds a factor of a few. Quite generally, substantial radiation drag leads to a faster deceleration, strengthening of the reverse shock and weakening of the forward shock. In the self-similar model with $n=1$ the 4-velocity of the upstream flow with respect to the shock increases from $u^\prime_u\simeq 0.5$ for a uniform ambient density profile and no radiation ($\alpha=0$), to about $u^\prime_u\simeq 2.1$ in the asymptotic limit of maximum efficiency (obtained for $\alpha >50$). The Lorentz factor in the asymptotic limit evolves roughly as $\Gamma\propto t^{-10}$.  

For numerical reasons the simulations were limited to modest values of $\alpha$, but show similar trends. The numerical experiments enabled us to follow the evolution of the system from the onset of fluid collision, and demonstrate the gradual strengthening of the reverse shock (and weakening of the forward shock) over time. For convenience we adopted intensity profile (of the seed radiation) that scales as $r^{-1}$, which is close to the flat profile obtained for blazars from detailed calculations.  Since for a sufficiently large drag the deceleration scale is much smaller than the shock radius, the details of the intensity profile has little effect on the evolution.    For our choice of constant drag coefficient, the deceleration continues indefinitely.   In reality the drag will be strongly suppressed once the Lorentz factor of the emitting (shocked) fluid drops to values at which our "beaming approximation" breaks down.   In this regime strong cooling still ensues, but with little momentum losses.  Our late time results for the cases $\alpha=10$ and $\alpha=15$ are therefore not reliable.  More detailed calculations of the drag terms are needed to follow the evolution in the mildly relativistic regime.

The effect of radiation drag on the dynamics of the shock might have important implications for the resultant emission. On the one hand, the strengthening of the reverse shock leads to enhanced efficiency of internal shocks, in particular under conditions at which the reverse shock is sub-relativistic in the absence of external radiation.  On the other hand, the deceleration of the shocked fluid, that emits the observed radiation, leads to a dramatic change in the beaming factor. This temporal change in beaming factor can significantly alter the light curves, particularly for observers viewing the source off-axis.  Detailed calculations of variable emission from dragged shocks is beyond the scope of this paper, but our analysis suggests that detailed emission models, at least in blazars, should account for such effects.

\begin{acknowledgments}
This research was supported by  a grant from the Israel Science Foundation no. 1277/13.
\end{acknowledgments}


\providecommand{\noopsort}[1]{}\providecommand{\singleletter}[1]{#1}%

\end{document}